\documentclass[sigconf]{acmart}

\usepackage{stfloats}

\begin{document}

\title{Clustering MOOC Programming Solutions to Diversify Their Presentation to Students}

\author{Elizaveta Artser}
\affiliation{
  \institution{\textit{JetBrains Research}}
  \city{Munich}
  \country{Germany}
}
\email{elizaveta.artser@jetbrains.com}

\author{Anastasiia Birillo}
\affiliation{
  \institution{\textit{JetBrains Research}}
  \city{Belgrade}
  \country{Serbia}
}
\email{anastasia.birillo@jetbrains.com}

\author{Yaroslav Golubev}
\affiliation{
  \institution{\textit{JetBrains Research}}
  \city{Belgrade}
  \country{Serbia}
}
\email{yaroslav.golubev@jetbrains.com}

\author{Maria Tigina}
\affiliation{
  \institution{\textit{JetBrains Research}}
  \city{Amsterdam}
  \country{Netherlands}
}
\email{maria.tigina@jetbrains.com}

\author{Hieke Keuning}
\affiliation{
  \institution{\textit{Utrecht University}}
  \city{Utrecht}
  \country{Netherlands}
}
\email{h.w.keuning@uu.nl}

\author{Nikolay Vyahhi}
\affiliation{
  \institution{\textit{Hyperskill}}
  \city{Boston}
  \country{United States}
}
\email{nikolay.vyahhi@hyperskill.org}

\author{Timofey Bryksin}
\affiliation{
  \institution{\textit{JetBrains Research}}
  \city{Limassol}
  \country{Cyprus}
}
\email{timofey.bryksin@jetbrains.com}

\renewcommand{\shortauthors}{Artser, Birillo, Golubev, Tigina, et al.}

\begin{abstract}
In many MOOCs, whenever a student completes a programming task, they can see previous solutions of other students to find potentially different ways of solving the problem and to learn new coding constructs. However, a lot of MOOCs simply show the most recent solutions, disregarding their diversity or quality, and thus hindering the students' opportunity to learn.

In this work, we explore this novel problem for the first time. To solve it, we adapted the existing plagiarism detection tool JPlag to Python submissions on Hyperskill, a popular MOOC platform. However, due to the tool's inner algorithm, JPLag fully processed only 46 out of 867 studied tasks. Therefore, we developed our own tool called \textsc{Rhubarb}. This tool first standardizes solutions that are algorithmically the same, then calculates the structure-aware edit distance between them, and then applies clustering. Finally, it selects one example from each of the largest clusters, thus ensuring their diversity. \textsc{Rhubarb} was able to handle all 867 tasks successfully. 

We compared different approaches on a set of 59 real-life tasks that both tools could process. Eight experts rated the selected solutions based on diversity, code quality, and usefulness. The default platform approach of simply selecting recent submissions received on average 3.12 out of 5, JPlag — 3.77, \textsc{Rhubarb} — 3.50. To ensure both quality and coverage, we created a system that combines both tools. We conclude our work by discussing the future of this new problem and the research needed to solve it better.

\end{abstract}



% Min 3 keywords

\maketitle
\vspace{0.5cm}

\section{Introduction}\label{sec:introduction}

Massive open online courses (MOOCs) are becoming more and more popular nowadays, serving the ever-growing need for accessible education~\cite{oh2020design}. 
Programming MOOCs usually consist of theoretical and practical parts that allow students to learn coding concepts step by step~\cite{staubitz2015towards}. 
The correctness of practical coding tasks is typically validated by automated testing (assessment) systems~\cite{staubitz2015towards}, however, because of the large number of students, it is difficult to control other aspects of writing code~\cite{kinash2013moocing}, \textit{e.g.}, code quality or specific ways the task can be solved. At the same time, these are very important for students' development~\cite{tigina2023analyzing, keuning2019teachers, renkl2014toward, brusilovsky2001webex}.

In some cases, teachers may prepare several different approaches to solve every task, however, this is not always feasible. One other way of learning these different approaches is sharing knowledge among students~\cite{kavitha2015knowledge, liu2014knowledge, sein2016students}. This is currently being actively studied and is even already implemented in several popular MOOCs, \textit{e.g.}, Codewars~\cite{codewars} or Hyperskill~\cite{jetbrainsAcademy}. These platforms allow students to see the solutions of other users after they solve the given task. The main problem here is that these solutions might not be diverse enough, since the majority of them are very similar. 

For this reason, it is crucial to cut the number of possible solutions to show only different ones to the student. As far as we know, such a problem has not yet been actively studied from a \textit{student} perspective, but a lot of tools were developed to solve similar problems from the perspective of a \textit{teacher} in related research areas of plagiarism detection~\cite{prechelt2000jplag, schleimer2003winnowing} and feedback generation~\cite{glassman2015overcode, gulwani2018automated}. 
Specifically, several tools like JPlag~\cite{prechelt2000jplag} and MOSS~\cite{schleimer2003winnowing} were developed to detect plagiarism in student submissions and can cluster them according to a plagiarism level. 

In this work, we strived to explore this novel problem not only for research purposes, but also to improve this functionality for Python tasks on Hyperskill~\cite{jetbrainsAcademy}, a large MOOC platform where the current approach is to simply show the latest solutions. First of all, we tried to adapt the existing approaches to this task. We chose JPlag~\cite{prechelt2000jplag} due to its simplicity, maturity, and popularity, and adapted it to be able to run for our purposes in the clustering mode. Like in many other popular tools, the algorithm~\cite{wise1993string} that JPlag uses to compare code snippets cannot handle short programs. This is an especially important disadvantage for Python solutions, since they can be short even for complex tasks. 

To solve the defined problem for solutions of any size, we also developed our own tool called \textsc{Rhubarb}~\cite{rhubarb}. Firstly, we developed a library~\cite{bumblebee} with a set of 12 Python transformations for standardizing code, \textit{e.g.}, anonymizing names of variables, removing dead code, etc. This helps to ignore syntactic changes that do not affect the overall idea of the solution. Next, \textsc{Rhubarb} calculates the edit distance between the standardized solutions using a state-of-the art tool called GumTree~\cite{falleri2014fine, teles2023code, mafi2024regression}, which takes into account the code structure. Then, using the calculated distances, it applies hierarchical agglomerative clustering~\cite{lukasova1979hierarchical}. Finally, the tool selects one example from each of the largest clusters, also considering the code quality of the solutions using Hyperstyle~\cite{birillo2022hyperstyle}, a tool already embedded into the Hyperskill platform.

The platform's team provided us with a dataset of 867 Python tasks with different levels of complexity and a total of 305,584 solutions to them. Of these tasks, JPlag was only able to fully process 46 (5.3\%), with 434 more (50.1\%) being processed partially (some solutions being skipped by the tool) and 387 (44.6\%) not processed at all, including some complex tasks. On the other hand, \textsc{Rhubarb} successfully processed 100\% of the data.

Finally, for the pilot evaluation of different approaches, we used 59 tasks, the ones for which JPlag was able to process at least 90\% of the solutions. For each of them, three groups of five solutions were generated --- \textbf{(1)}~with the default platform approach of selecting recent submissions, \textbf{(2)} with JPlag, and \textbf{(3)} with \textsc{Rhubarb}. Then, eight experts rated these groups on a Likert scale from the standpoint of diversity, code quality, and usefulness, with an average of 2.3 experts per task. The resulting average scores are 3.12 out of 5 for the existing platform approach of simply selecting the most recent solutions, 3.77 for JPlag, and 3.50 for \textsc{Rhubarb}, so both tools performed better than the default approach from Hyperskill. Since in the real MOOC, processing all solutions is crucial, and the difference in the quality between JPlag and our approach is not drastic, we implemented a system that employs JPlag on the 5.3\% of tasks it can fully process, and \textsc{Rhubarb}~ on the remaining 94.7\%. 

You can find the source code of \textsc{Rhubarb} in our GitHub repository~\cite{rhubarb}. Separately, you can find a library with standardizing transformations for Python~\cite{bumblebee} that can be useful in other tasks. Finally, the supplementary materials for the paper are available on Zenodo~\cite{artifacts}.

Overall, the contributions of this paper are the following:

\begin{itemize}
    \item \textbf{\textsc{Rhubarb}}, a tailored clustering tool for selecting diverse Python solutions for showing in MOOCs. 
    \item As a part of \textsc{Rhubarb}, a library of \textbf{code transformations} for Python that can be useful in other applications.
    \item A \textbf{pilot evaluation} of the concept and \textsc{Rhubarb} with eight experts on 59 real tasks from Hyperskill, a large MOOC platform, showing that it outperforms the default platform approach, as well as a discussion of the results together with the potential future of the proposed problem and the solution.
\end{itemize}

\section{Motivating Example}\label{sec:motivating:example}

\begin{figure*}[t]
    \centering
    \includegraphics[width=\linewidth]{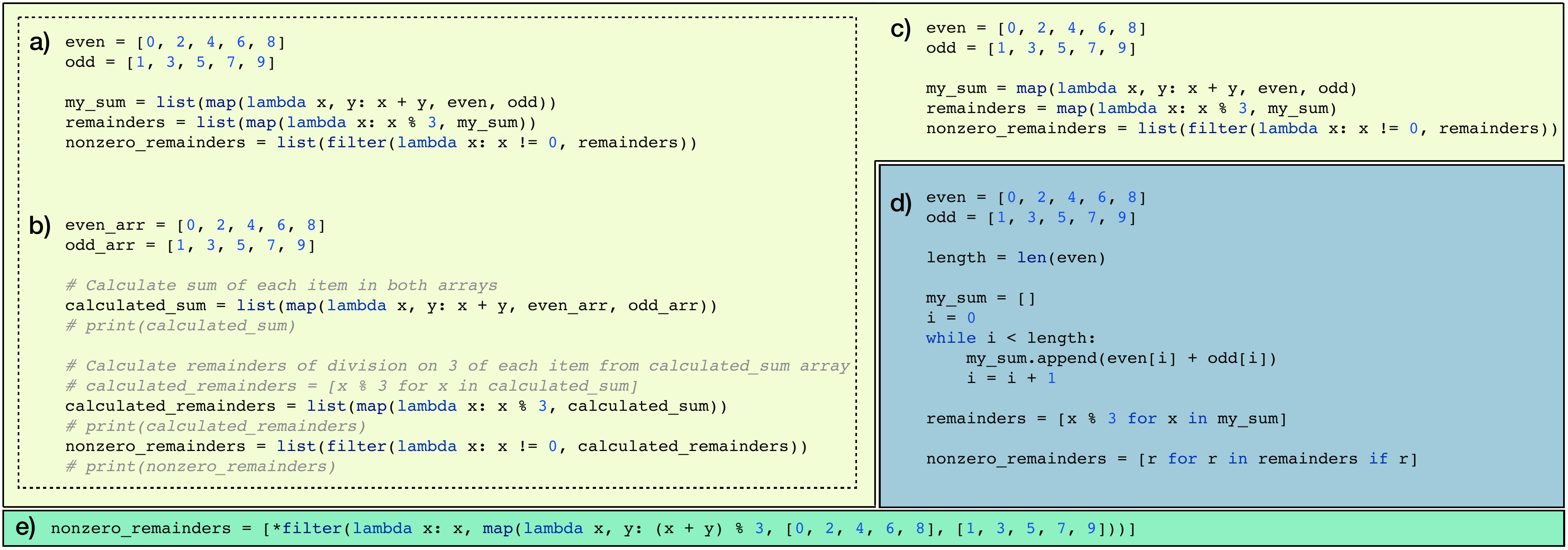}
    \caption{Examples of different students' approaches to solving the same task. Solutions (a) and (b) are algorithmically the same, while solution (c) is not, but still very similar to them. On the other hand, solutions (d) and (e) are rather different.}
    \Description{Examples of different students' approaches to solving the same task. Solutions (a) and (b) are algorithmically the same, while solution (c) is not, but still very similar to them. On the other hand, solutions (d) and (e) are rather different.}
    \label{fig:examples}
\end{figure*}

In Figure~\ref{fig:examples}, we can see five different solutions to the same task about number remainders and functional programming in Python. Let us imagine that a student submitted solution \textbf{(a)} and we want to show them some other, different ones.

Firstly, we notice that solution \textbf{(b)} is not just similar, it's \textit{algorithmically the same}. In this study, we define two code fragments to be algorithmically the same if they can be obtained from one another using some common transformations that do not affect the logic and can be applied without understanding the code, \textit{e.g.}, renaming variables, adding comments, etc. If these two solutions were to be brought to a standard form, they would be the same, thus we do not even need to consider solution \textbf{(b)} for showing to the student. 

Solution \textbf{(c)} is a different case. It looks even more similar to \textbf{(a)}, to the degree where it may take some time to find the difference, and so it also probably should not be shown to the student. However, it is actually \textit{algorithmically different}, since it skips using the \texttt{list} function twice. This does not matter in this particular context (\texttt{map} produces an iterator~\cite{pythonMap} that works for this task), but this is not always the case, and this change could alter the logic in different contexts. To take such cases into account, it is necessary to cluster similar solutions, since transformations will not cover them.

On the other hand, solutions \textbf{(d)} and \textbf{(e)} are very different both compared to previous solutions and to each other. Solution \textbf{(d)} uses an iterative approach, while solution \textbf{(e)} is an impressive but difficult to read one-liner. From the standpoint of diversity, however, they are relevant and interesting to be shown to the discussed student.

This example demonstrates several key points that influence our study: \textbf{(1)} we can use code transformations to standardize code snippets and merge them (\textit{e.g.}, \textbf{(a)} and \textbf{(b)}); \textbf{(2)} some minor changes can not be considered by standardization, but should be handled by a clustering algorithm (\textit{e.g.} \textbf{(c)}); \textbf{(3)} the shortness of the solution does not always correlate with task complexity (\textit{e.g.}, \textbf{(e)}), and short submissions should also be handled in our context.

\section{Background}\label{sec:background}

The topic of clustering student submissions is being actively studied. Some studies are aimed at finding plagiarism in student solutions~\cite{prechelt2000jplag, schleimer2003winnowing}, while others aim to show their variety: to help teachers provide feedback on programming assignments~\cite{gulwani2018automated, glassman2015overcode}, or generate human-readable descriptions to help teachers identify different approaches~\cite{effenberger2021interpretable}. 

JPlag~\cite{prechelt2000jplag} is a popular plagiarism detection tool that uses a greedy string tiling algorithm~\cite{wise1993string} to find the distance between solutions. During each pair-wise comparison, JPlag attempts to cover one string with substrings taken from the other as well as possible, and then applies the Dice coefficient~\cite{dice1945measures} to the results to obtain the final similarity score. JPlag also takes into account the programming language keywords within this algorithm. The tool works for a lot of popular languages, including all the major ones. It also has a special mode for clustering solutions, where each cluster contains solutions that the tool considers to be plagiarized. The main disadvantage of JPlag is the restriction on the minimum number of tokens in the code fragment, which does not allow working with short submissions. This is important in the context of our work, since in Python even complex tasks can have short solutions, \textit{e.g.}, solution \textbf{(e)} from Figure~\ref{fig:examples}. In addition, this approach considers code snippets as a set of string tokens, which makes it impossible to analyze the code structure and the used constructs. 

Another popular plagiarism detection tool is MOSS~\cite{schleimer2003winnowing}, a Web service that also uses an adaptation of the greedy string tiling algorithm~\cite{wise1993string}. The key difference is that the fragments of code marked as similar appear in no more than \textit{N} submissions. MOSS also allows for the exclusion of code that is directly repeated between solutions, such as the pre-written templates provided by the task's creator. The tool supports many languages, however, the source code of the tool is closed and cannot be reused for research purposes, including ours. 

OverCode~\cite{glassman2015overcode} helps teachers provide appropriate feedback to students at scale. This tool reformats the code using abstract syntax tree (AST) analysis, but only in terms of style, removing comments and renaming variables; the comparison itself is still done by string-matching its lines. OverCode was developed for Python 2, did not update since 2016, and fails on fragments with newer constructs, \textit{e.g.}, solution \textbf{(e)} from Figure~\ref{fig:examples}. 

CLARA~\cite{gulwani2018automated} clusters existing correct student solutions, and uses these clusters to generate a possible repair for a new incorrect solution. To compare submissions, this tool builds control-flow graphs (CFGs) and compares them dynamically via test inputs and variable values on each program step. The tool supports C and Python, however, it only supports a limited number of language constructs, which complicates running the tool on complex tasks. Also, matching CFGs is in general an NP-hard problem, which makes it impossible to run CLARA on a big MOOC platform.

Finally, a recent study~\cite{effenberger2021interpretable} introduced the concept of \textit{interpretable} clustering for Python. The paper aimed to develop a clustering algorithm for applications where interpretability is key, \textit{e.g.}, to show different ways to solve a task to the teacher. In this approach, each cluster has a short description to indicate the key features of its submissions. The clustering algorithm itself extracts different language constructs from submissions, and then uses them to generate frequent patterns. This approach looks promising, but the current rules for extracting constructs are designed only for simple tasks, and the source code for this approach is not available.

Overall, it can be seen that of the mentioned tools, only JPlag can be adapted to solve our problem at scale.

\section{JPlag}

Firstly, we decided to adapt the plagiarism detection tool JPlag to our task. This tool uses a greedy string tiling algorithm to calculate the distance between students' submissions and builds a clustered graph, where each cluster represents solutions that it considers to be plagiarized. The solution's code is represented as a list of tokens, which are obtained by a special language-specific parser.

We ran JPlag from command line by giving it the path to input and output directories, as well as the settings of the clustering: its type (agglomerative hierarchical clustering with complete linkage) and a threshold. We used this type of clustering, because it is understandable and interpretable~\cite{sembiring2011comparative}. The fine-tuning of the threshold is described in Section~\ref{sec:validation:thresholds}. JPlag calculates the similarity between each pair of solutions using the Dice coefficient~\cite{dice1945measures}, which measures how similar two sets of tokens are, and then clusters them using the provided threshold.

The output consists of a list of clusters and a list of solutions for each of them. From the output, we select five of the largest clusters, and from each of them, simply pick the first solution as presented, which comes from the first clustered pair, the one with the highest Dice coefficient.

However, JPlag cannot properly process all the submissions. Firstly, solutions are ignored if they have fewer than the specified number of particular tokens (12 for Python), and we cannot simply decrease this value because it is a base requirement for greedy string tiling algorithms. Secondly, only a limited number of tokens is supported that can be used for the clustering itself. As a result, for the example in  Figure~\ref{fig:examples}, JPlag was only able to connect solutions \textbf{(a)} and \textbf{(b)}, only partially parsed solutions \textbf{(c)} and \textbf{(d)}, and entirely skipped solution \textbf{(e)} because of its size. Overall, in the provided platform dataset, JPlag could process all solutions for only 5.3\% of tasks. More detailed statistics about this are provided in Section~\ref{sec:data}.

\section{Rhubarb}

To overcome the shortcomings of JPlag so that all the submissions on the Hyperskill platform could be processed, we developed our own tool called \textsc{Rhubarb}. Its overview is presented in Figure~\ref{fig:clustering:pipeline}, below we describe each stage in detail.

\subsection{Standardization of Submissions}\label{sec:clustering:unification}

\begin{figure}[t]
    \centering
    \includegraphics[width=1\linewidth]{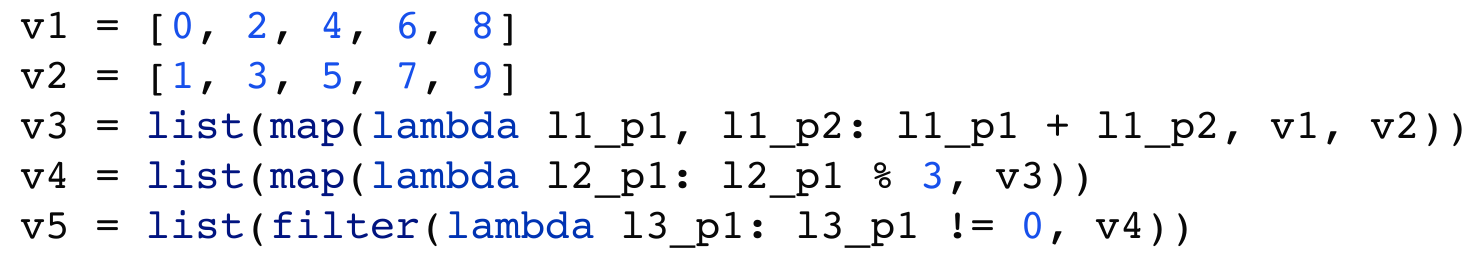}
    \caption{The standard form of solutions \textbf{(a)} and \textbf{(b)} from Figure~\ref{fig:examples}.}
    \Description{The standard form of solutions \textbf{(a)} and \textbf{(b)} from Figure~\ref{fig:examples}.}
    \label{fig:clustering:unification}
\end{figure}

In Section~\ref{sec:motivating:example} and Figure~\ref{fig:examples}, we showed an example of code fragments that are algorithmically the same. The goal of the first step is to bring such solutions to a standard form~\cite{xu2003transformation, glassman2015overcode}. The standard form for solutions \textbf{(a)} and \textbf{(b)} from Figure~\ref{fig:examples} is shown in Figure~\ref{fig:clustering:unification}. This is achieved using \textit{code transformations}.

\begin{figure*}[t]
    \centering
    \includegraphics[width=\linewidth]{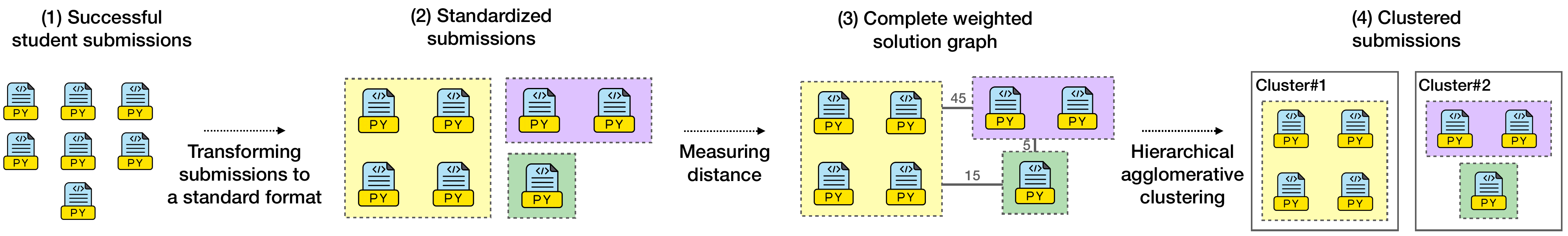}
    \caption{The general overview of \textsc{Rhubarb}.}
    \Description{The general overview of \textsc{Rhubarb}.}
    \label{fig:clustering:pipeline}
\end{figure*}

Our implementation of code transformations relies on the IntelliJ Platform~\cite{kurbatova2021intellij}, which powers IntelliJ-based IDEs. Specifically, we employ the Program Structure Interface (PSI), a layer of the IntelliJ Platform responsible for parsing files and creating the syntactic and semantic code model. A PSI file consists of a hierarchy of PSI elements, which enable the exploration of the internal structure of the source code as the IntelliJ Platform interprets it. PSI is similar to AST, however, while AST serves as a general interface for working with tree nodes, PSI is a more specialized implementation with specific language-related functionalities. In particular, it is much more tailored for changing the code, which is why we chose it.

We developed a library, in which several transformations for PSI are implemented. In total, there are 12 transformations for Python, the major ones are:

\begin{itemize}
    \item \textbf{Anonymization} --- changing all the identifiers in the program to generic ones.
    \item Removing \textbf{comments}, \textbf{empty lines}, and \textbf{dead code}.
    \item Standardizing \textbf{assignments} (from \texttt{x += 1} to \texttt{x = x + 1}) and \textbf{equations} (from \texttt{x < y} to \texttt{y > x}, always ``greater'').
    \item \textbf{Propagating constants} (from \texttt{1 + 2 + 3} to \texttt{6}).
\end{itemize}

The library can be found separately on GitHub~\cite{bumblebee}. We developed this set of transformations according to an existing study~\cite{rivers2017data} that described standardizing transformations for constructs commonly found in student submissions. 
We expect that the library can be useful for other potential tasks, \textit{e.g.}, hint generation, clone detection, or verifying program correctness
without using test cases.
The library is written in Kotlin and can be used directly or via an API. 

After the transformations are applied, the standardized solutions are saved in a so-called solution graph (see \textbf{(2)} in Figure~\ref{fig:clustering:pipeline}). At the standardization stage, the solution graph does not contain any edges. The vertices consist of a unique ID, standardized code, and a list of solution IDs that correspond to it. Although standardization in this form already constitutes clustering of code solutions, it only combines solutions with exactly the same semantics. Solutions with similar approaches (\textit{e.g.}, \textbf{(c)} from Figure~\ref{fig:examples}) may still be in different groups, hence the need for subsequent clustering.

\subsection{Measuring Distance}\label{sec:clustering:distance}

For clustering solutions, it is necessary to introduce a particular distance metric between them. The chosen metric is the code \textit{edit distance} in terms of AST nodes, which is similar to text-based approaches but considers code structure. If one solution can be transformed into another with a few edit operations ({\it i.e.}, the edit distance between them is small), these solutions are likely to use similar approaches and should be in the same cluster.

The code edit distance is calculated using GumTree~\cite{falleri2014fine}, a state-of-the-art tool for comparing and analyzing code. GumTree is actively used in similar recent works~\cite{teles2023code} and remains one of the best code differencing tools~\cite{mafi2024regression}. In GumTree, the edit distance measures the number of editing operations (insertions, deletions, and moves of AST nodes) needed to transform one code version into another. Since some nodes in ASTs constitute entire code blocks, we ``expand'' their inner nodes and consider their sizes when calculating distances. Compared to other methods of calculating the edit distance, GumTree uses the analysis of syntax trees and the calculation of paths between tree nodes. As the tool uses its own internal format for trees, we implemented a custom converter from PSI to the GumTree tree to call its API directly.

The distance measuring process involves turning the previously obtained solution graph into a complete weighted graph (see \textbf{(3)} in Figure~\ref{fig:clustering:pipeline}). For each pair of standardized solutions, the distance is calculated and added as an edge with a weight. Consider the code snippets from Figure~\ref{fig:examples}. Code snippets \textbf{(a)} and \textbf{(b)} can be considered as one node \textbf{(a-b)}, since they merge during the standardization step. The distance between nodes \textbf{(a-b)} and \textbf{(c)} is $56$, while the distance between \textbf{(d)} and \textbf{(e)} is $530$ (see Figure~\ref{fig:clustering:distance_limit}), indicating that \textbf{(a-b)} and \textbf{(c)} are more similar to each other.

\subsection{Hierarchical Agglomerative Clustering}\label{sec:clustering:clustering}

The next step of the algorithm is clustering the standardized solutions based on the calculated distances between them. Similarly to the JPlag pipeline, we chose the hierarchical agglomerative clustering (HAC)~\cite{anderberg2014cluster} as the basic clustering algorithm because of its popularity and interpretability~\cite{sembiring2011comparative}. It iteratively merges the most similar pairs of data points or clusters until a stopping criterion is met. A clear stopping criterion for our setting is the restriction on the distance between solutions within a cluster. In this case, the distance between two standardized solutions within one cluster cannot exceed the pre-set distance limit at any point in the HAC algorithm. This hyperparameter can be set up through the tool's arguments. The complete linkage~\cite{anderberg2014cluster} metric of inter-cluster distance maintains this invariant: the distance between two clusters is given by the maximum weight edge between points in the two clusters.

\begin{figure}[t]
    \centering
    \includegraphics[width=0.4\textwidth]{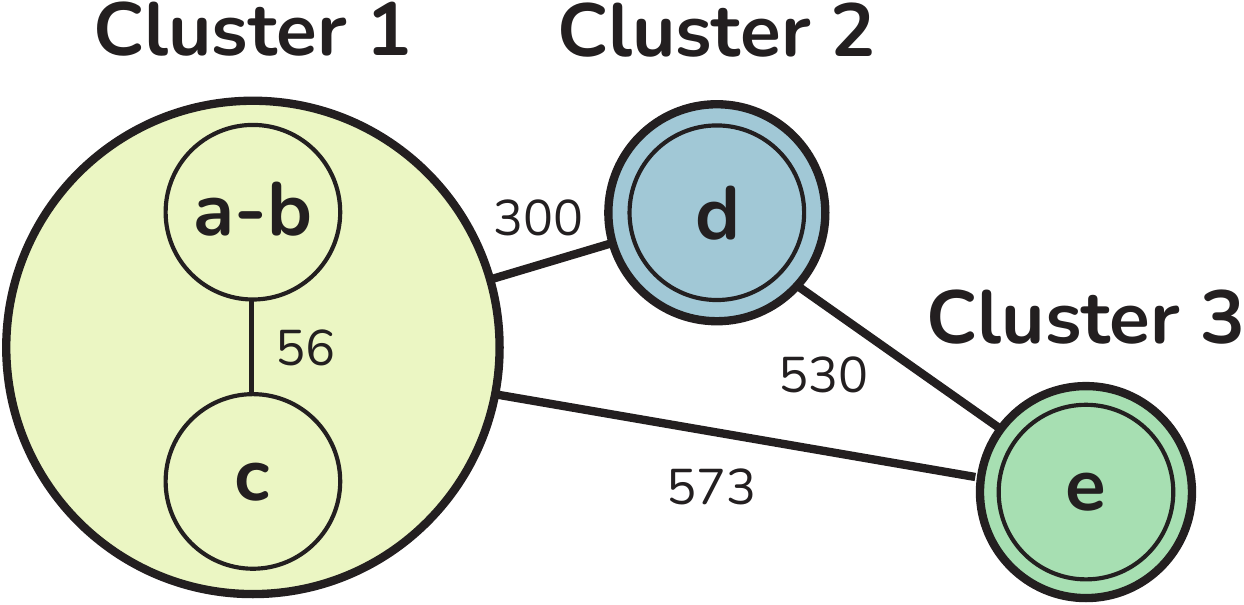}
    \caption{The graph with distances between the standardized solutions and clusters for solutions from Figure~\ref{fig:examples}.}
    \Description{The graph with distances between the standardized solutions and clusters for solutions from Figure~\ref{fig:examples}.}
    \label{fig:clustering:distance_limit}
\end{figure}

We also decided to supplement HAC with a heuristic that relates to the considered context. When the distance between pairs of clusters is equal, we first merge clusters of smaller size. Merging small clusters first can help identify rare approaches to solving a problem that may be lost if included in larger clusters.

The result of this step is another complete weighted graph (see \textbf{(4)} in Figure~\ref{fig:clustering:pipeline}). Its vertices are clusters containing a unique cluster ID and a list of standardized solutions. The edge weights of the cluster graph are the inter-cluster distances calculated during clustering. Each cluster is converted from a list of standardized solutions to a list of original student solutions from the platform and saved. An example of a simple graph that corresponds to the examples from Figure~\ref{fig:examples} is shown in Figure~\ref{fig:clustering:distance_limit}, with both inter-solution and inter-cluster distances. As we can see, code snippets \textbf{(a)}, \textbf{(b)}, and \textbf{(c)} represent the same cluster, and the submissions \textbf{(d)}, and \textbf{(e)} build their own clusters, just as we discussed.

\subsection{Selecting Representative Solutions}\label{sec:clustering:output:data}

Finally, we need to select representative examples to present them to the user. Similar to JPlag's pipeline, we take the five largest clusters, however, instead of selecting the first solutions within them, we additionally sort them by code quality. Specifically, we take the code quality grade provided by the Hyperstyle tool~\cite{birillo2022hyperstyle} that is used in the studied Hyperskill platform. We then sort the solutions in the cluster based on this grade, and select the example from the ones with the highest grade. This allows us to additionally ensure that the provided examples are of good quality. In principle, any other sorting can be used in \textsc{Rhubarb}, according to the needs of a particular platform, \textit{e.g.}, code quality, size, etc.

\subsection{Limitations}

The developed tool has several limitations. First, standardization can take a long time, since we need to apply 12 transformations, and building and resolving a PSI tree is a complex process. We already implemented the ability to serialize the result of this step to reuse the standardized submissions, and we plan to add parallelization in future work. Next, to cluster student submissions, a complete weighted graph is built. This can be time- and memory-consuming for a large number of submissions. To overcome this drawback, we plan to support incremental graph recalculation.

\section{Evaluation}
\subsection{Dataset}\label{sec:data}

To fine-tune and evaluate our approach, we used the data provided to us by the Hyperskill platform~\cite{jetbrainsAcademy}. The platform's policy allows to use anonymized data for research purposes, and the data did not include any personal or identifying information. The platform has Hyperstyle~\cite{birillo2022hyperstyle} as the embedded code style tool, so the code quality grade is already included in the data. The dataset represents all correct Python solutions submitted within $16$ weeks, a total of 305,584 solutions for 867 different coding tasks. The data contains the ID and the code of each submission, its code quality, and a timestamp. The timestamp helps to simulate the default platform's approach and extract only the several last solutions.

A preliminary run on all tasks showed that JPlag failed to process 387 tasks (44.6\%). For 298 of them, the report file was not generated, and for another 89, it generated an empty set of clusters. 
Next, 434 tasks (50.1\%) were processed partially, {\it i.e.}, some solutions were absent in the report file. Finally, 46 tasks (5.3\%) were processed in full, {\it i.e.}, all solutions were present in the report. The detailed distribution of the processed tasks can be found in Table~\ref{table:jplg_fail}. Such a drastic skipping of data indicates the importance and the necessity of \textsc{Rhubarb} for the practical use on the platform.

\begin{table}[t]
\centering\tabcolsep4pt
\caption{\label{table:jplg_fail} The absolute and relative number of tasks, for which JPlag can process different percentage of solutions. \textit{E.g}, the column (10-50] indicates the number of tasks, for which 10-50\% of all submissions were present in JPlag's output.}
\Description{The absolute and relative number of tasks, for which JPlag can process different percentage of solutions. \textit{E.g}, the column (10-50] indicates the number of tasks, for which 10-50\% of all submissions were present in JPlag's output.}
\begin{tabular}{c c c c c c c}
\toprule
       & \textbf{0}   & \textbf{(0-10]} & \textbf{(10-50]} & \textbf{(50-90]} & \textbf{(90-100]} & \textbf{100} \\\midrule
\textbf{№ of tasks} & 387 & 134    & 79       & 109    & 112      & 46 \\
\textbf{\% of tasks} & 44.6 & 15.5 & 9.1 & 12.6 & 12.9 & 5.3\\
\bottomrule
\end{tabular}
\end{table}

For a fair comparison with enough data, we decided to consider the tasks, for which JPlag could process at least 90\% of solutions (the last two columns in Table~\ref{table:jplg_fail}). There are a total of 158 such tasks, among which 76 have at least 1,000 solutions. These 76 (8.8\%) tasks were divided into two groups for two different purposes: 17 tasks were used for fine-tuning the clustering hyperparameters, and 59 were used for the final evaluation and comparison.

\subsection{Fine-tuning Parameters}\label{sec:validation:thresholds}

Both JPlag and \textsc{Rhubarb} have parameters that have to be set --- a threshold for JPlag and the distance limit for \textsc{Rhubarb}. Before conducting the comparison of these algorithms against the default platform approach, it is necessary to select these parameters. We decided to carry out both procedures (fine-tuning and comparison) in a similar fashion, by conducting manual labeling. Since this paper explores a novel application of such tools, the fine-tuning is also largely exploratory, and we leave a more detailed evaluation of parameters with more experts for future work.

For the fine-tuning, we took the 17 tasks as described in the previous section, and for each of them, obtained 5 different solutions for showing using either JPlag or \textsc{Rhubarb} with different parameters, respectively. For example, for JPlag, we compared four different values of the threshold, so for each of the 17 tasks, we obtained four different groups of five solutions and rated them. The rating was carried out on the 1-to-5 scale, with \textit{1 indicating  the least suitable sample in terms of the diversity, code quality, and usefulness} and \textit{5 indicating the most suitable sample in terms of the diversity, code quality, and usefulness}. A single score was given, reflecting an overall assessment of all three dimensions combined.
The fine-tuning of the parameters was carried out by the first two authors of the paper, who have several years of experience working with MOOCs and programming in Python, and who labeled all 17 tasks. The final score for each setting was averaged between all 17 tasks and both labelers.

\begin{table*}[b]
\centering
\caption{\label{table:parval_jplag} The results of fine-tuning for JPlag threshold and \textsc{Rhubarb} distance limit.}
\Description{The results of fine-tuning for JPlag threshold and \textsc{Rhubarb} distance limit.}
\begin{tabular}{lcccclcccccccc}
\toprule
\multicolumn{5}{c}{\textbf{JPlag}} & \multicolumn{9}{c}{\textbf{\textsc{Rhubarb}}}\\
\cmidrule(lr){1-5}\cmidrule(lr){6-14}
\textbf{Threshold} & \textbf{0.2} & 0.4 & 0.6 & 0.8 & \textbf{Dist. limit} & 15 & 30 & 45 & \textbf{60} & 75 & 90 & 105 & 120 \\\cmidrule(lr){1-5}\cmidrule(lr){6-14}
\textbf{Avg. score} & \textbf{3.85} & 3.82 & 3.82 & 3.71 & \textbf{Avg. score} & 3.15 & 3.06 & 3.06 & \textbf{3.24} & 3.09 & 3.03 & 2.82 & 2.74 \\
\bottomrule
\end{tabular}
\end{table*}

The results of fine-tuning are presented in Table~\ref{table:parval_jplag}. For JPlag, we compared four values of the threshold: $0.2$, $0.4$, $0.6$, and $0.8$. It can be seen that the results are very similar; we did not notice particular differences in the output. We selected $0.2$ as the final threshold as it is marginally better, and also because it is the default threshold.

For \textsc{Rhubarb}, we evaluated distance limits from $15$ to $120$, a range that showed meaningful results in preliminary experiments, with a step of $15$, a total of eight values. Here, the results showed a little more difference, although only at the extreme ends of the range. We selected the distance limit of $60$, which received the higher average score. The selected thresholds need to be further evaluated with more experts, which we leave for future work.

\subsection{Comparison}\label{sec:validation:approach}

Having selected the best parameters for both JPlag and \textsc{Rhubarb}, we ran the main evaluation to compare them to each other and to the current default approach of the Hyperskill platform that simply selects the latest solutions. The format of the evaluation was the same as for the fine-tuning. The final labelling was carried out by the first author and seven other invited experts. Among these experts were experienced software developers, with four having more than 5 years of programming experience, and computer science teachers, with two having more than 5 years of teaching CS courses. For a total of 59 tasks, there was an average 2.3 labelers per task.  The outputs of the three approaches were always shuffled in their order and the approaches' names were hidden. The final scores were first averaged between experts for each task, and then averaged between all tasks for each approach.

The averaged results are: \textit{Platform} -- 3.12, \textbf{\textit{JPlag} -- 3.77}, \textit{\textsc{Rhubarb}} -- 3.50. As evident from the obtained data, both JPlag and \textsc{Rhubarb} offer more diverse solutions than those currently displayed on the Hyperskill platform. This indicates the usefulness of clustering in this context and of developing a new approach for this task. 

On the one hand, the evaluation has shown that JPlag works better than \textsc{Rhubarb}, while both of them work better than the default approach. On the other hand, the comparison itself was carried on just a small sample of the data, since JPlag fails for the majority. The \textsc{Rhubarb} tool handled 100\% of the dataset and can produce clusters with better quality than the default platform's solution for all tasks. Importantly, JPlag can not be extended to all cases, since greedy string tiling algorithms by their definition can not be used for short submissions, while the quality of \textsc{Rhubarb} can be potentially improved, and it is already rather close to JPlag. For now, we designed a system that employs JPlag on the 5.3\% of tasks it can fully process, and uses \textsc{Rhubarb} on the remaining tasks.

\section{Discussion}

With this work, we aimed to explore the novel task of selecting diverse solutions to programming tasks for showing to students, both for research purposes and for using in a real MOOC platform. While we only conducted preliminary fine-tuning and pilot evaluation, some general findings are worth discussing.

\textbf{Comparison to the default approach}. Firstly and most obviously, it can be seen that both clustering approaches outperform the default platform approach of showing the latest solutions. This indicates the clear room for improvement for the existing implementation and the potential of the proposed concept in general. 

\textbf{Comparison to JPlag}. At the same time, among the clustering-based approaches, the existing tool JPlag demonstrated better performance than \textsc{Rhubarb}. While this does not make \textsc{Rhubarb} obsolete due to it being able to handle all the submissions, it does show the need to integrate other distance measuring and clustering techniques into our approach. In the future, different methods should be combined and evaluated for this task.

\textbf{Overall scores}. Finally, one could argue that all the obtained scores are average at best, raising more research questions about showing other students' solutions in the first place. We argue that this does not decrease the value of the proposed problem and our work in particular. Firstly, other ways of teaching different ways of solving problems (\textit{e.g.}, pre-written teachers' solutions) are not always feasible on MOOCs. Secondly, the system of showing other students' solutions is already implemented in several major MOOCs and thus already affects thousands of students. What this does indicate is the necessity to study this aspect of MOOCs more deeply, in particular, by comparing the proposed approaches in real life, by running A/B experiments with real students on MOOCs. 

We plan to do this in our future work, and we hope that this paper can serve as an important first step towards bringing the attention of other researchers to this aspect of MOOCs and its practical value.

\section{Conclusion and Future Work}\label{sec:conclusion}

This work introduces a new problem of selecting algorithmically diverse solutions in order to display them to students on MOOCs. To address it, we proposed a tool called \textsc{Rhubarb}~\cite{rhubarb} that brings Python solutions to a standard form, calculates edit distances between them via the GumTree tool, and selects final examples based on code quality. Using the data provided to us by the Hyperskill MOOC platform~\cite{jetbrainsAcademy}, we compared the default platform approach of simply showing the last student submissions, the adapted existing plagiarism tool JPlag, and \textsc{Rhubarb}. In the manual labelling by experts, the default approach received the average score of 3.12 out of 5, JPlag --- 3.77, and our tool --- 3.50. The comparison was performed only on a small sample of the data, since JPlag cannot fully process the majority of our tasks, with 44.6\% not being processed at all. In the end, we created a compound version that uses both tools together to handle 100\% of the data with good quality of the output that is higher than the standard platform's approach.

\textbf{Future work.} Our future work relates to both research and practice. From the research point of view, we could increase the quality of \textsc{Rhubarb} by improving the GumTree processing and trying different types of clustering, \textit{e.g.}, mean-shift algorithms for clustering~\cite{wu2007mean}. Also, it is imperative to conduct a more thorough evaluation of hyperparameters, and in general --- carry out more expert labeling. From the practical standpoint, it is critical to make sure that the system actually helps students, since in this work we only focused on the initial exploration of the problem. We plan to actually incorporate the developed system into the studied MOOC platform, since it outperforms the current version, and conduct studies into the effect of showing alternative solutions to students. In addition to validating our approach, this will allow to create a more general framework for exploring this problem, comparing different techniques, and collecting student feedback on other forms of diversifying their knowledge that might not be obvious.

\bibliographystyle{ACM-Reference-Format}
\bibliography{cite}

\end{document}